\documentclass[fleqn,10pt]{wlscirep}
\usepackage[utf8]{inputenc}
\usepackage[T1]{fontenc}

\title{The role of geography in the complex diffusion of innovations}

\author[a,b,c,d,*]{Balázs Lengyel}
\author[c,d]{Eszter Bokányi}
\author[a,e,f]{Riccardo Di Clemente} 
\author[g]{János Kertész}
\author[a,h,i]{Marta C. González}

\affil[a]{Massachusetts Institute of Technology, Department of Civil and Environmental Engineering, Cambridge MA, 02139, USA}
\affil[b]{International Business School Budapest, Budapest, 1037, Hungary}
\affil[c]{Agglomeration and Social Networks Lendület Research Group, Centre for Economic- and Regional Studies, Institute of Economics, Budapest, 1097, Hungary}
\affil[d]{Corvinus University of Budapest, Institute of Advanced Studies, Budapest, 1093, Hungary}
\affil[e]{University of Exeter, Computer Science Department, Exeter, EX4 4QF, United Kingdom}
\affil[f]{University College London, The Bartlett Centre for Advanced Spatial Analysis, London, WC1E 6BT, United Kingdom}
\affil[g]{Central European University, Department of Network and Data Science, Budapest, 1051, Hungary}
\affil[h]{University of California at Berkeley, Department of City and Regional Planning, Berkeley CA, 94720, USA}
\affil[i]{Energy Analysis and Environmental Impacts Division, Lawrence Berkeley National Laboratory, Berkeley Ca, 94720, USA}
\affil[*]{Corresponding address: lengyel.balazs@krtk.mta.hu}

\keywords{Social Networks $|$ Spatial Adoption $|$ Innovation $|$ Complex Contagion $|$ Diffusion Models} 

\begin{abstract}
The urban-rural divide is increasing in modern societies calling for geographical extensions of social influence modelling. Improved understanding of innovation diffusion across locations and through social connections can provide us with new insights into the spread of information, technological progress and economic development. In this work, we analyze the spatial adoption dynamics of iWiW, an Online Social Network (OSN) in Hungary and uncover empirical features about the spatial adoption in social networks. During its entire life cycle from 2002 to 2012, iWiW reached up to 300 million friendship ties of 3 million users. We find that the number of adopters as a function of town population follows a scaling law that reveals a strongly concentrated early adoption in large towns and a less concentrated late adoption. We also discover a strengthening distance decay of spread over the life-cycle indicating high fraction of distant diffusion in early stages but the dominance of local diffusion in late stages. The spreading process is modelled within the Bass diffusion framework that enables us to compare the differential equation version with an agent-based version of the model run on the empirical network. Although both model versions can capture the macro trend of adoption, they have limited capacity to describe the observed trends of urban scaling and distance decay.
We find, however that incorporating adoption thresholds, defined by the fraction of social connections that adopt a technology before the individual adopts, improves the network model fit to the urban scaling of early adopters. Controlling for the threshold distribution enables us to eliminate the bias induced by local network structure on predicting local adoption peaks. Finally, we show that geographical features such as distance from the innovation origin and town size influence prediction of adoption peak at local scales in all model specifications. 
\end{abstract}

\begin{document}

\flushbottom
\maketitle
\thispagestyle{empty}

\section*{Introduction}

Collective behavior, such as massive adoption of new  technologies , 
is a complex social contagion phenomenon \cite{centola2007complex}. Individuals are influenced both by media and by their social ties in their decision-making. This feature was first modelled in the 1960s with the Bass model of innovation diffusion \cite{bass1969new}. The model distinguishes between exogenous and peers’ influence and reproduces the observation that few early adopters are followed by a much larger number of early and late majority adopters, and finally, by few laggards \cite{rogers2010diffusion}. The differential equations of the Bass model have been extensively used to describe the diffusion process and forecast market size of new products and the time of their adoption peaks\cite{mahajan1991new}.

\vskip 0.1in
Only in the past two decades, the importance of the social network structure has become increasingly clear in the mechanism of peers’ influence \cite{centola2018behavior}. In spreading phenomena, individuals perform a certain action only when a sufficiently large fraction of their network contacts have performed it before \cite{schelling1978micromotives,granovetter1978threshold,valente1996social, watts2002simple, banerjee2013diffusion}. Complex contagion models, in which adoption depends on the ratio of the adopting neighbors, often referred to as adoption threshold \cite{centola2007complex, pastor2015epidemic}, have been efficiently applied to characterize the diffusion of online behavior \cite{centola2010spread} and online innovations \cite{karsai2016local,katona2011network}. In order to incorporate the role of social networks in technology adoption, the Bass model has been implemented through an agent-based model (ABM) version \cite{rand2011agent}. This approach is similar to other network diffusion approaches regarding the increasing pressure on the individual to adopt as network neighbors adopt; however, spontaneous adoption is also possible in the Bass ABM \cite{watts2007influentials}. The structure of social networks in diffusion, such as community or neighborhood structure of egos, are still topics of interest \cite{ugander2012structural,aral2017exercise}. Nevertheless, understanding how physical geography affects social contagion dynamics is still lacking \cite{centola2007complex}.

\vskip 0.1in
Early work on spatial diffusion has highlighted that adoption rate grows fast in large towns and in physical proximity to initial locations of adoption \cite{griliches1957hybrid, hagerstrand1968innovation}. It is argued that spatial diffusion resembles geolocated routing through social networks \cite{hagerstrand1968innovation}. Social contagion – similar to geolocated routing \cite{leskovec2014geospatial} – occurs initially between two large settlements located at long distances and then becomes more locally concentrated reaching smaller towns and short distance paths.
Facilitated by the observation of a large scale Online Social Network (OSN) over a decade, we capture for the first time these dynamics and provide insights into social network diffusion in its geographical space.

\vskip 0.1in
In this paper, we analyze the adoption dynamics of iWiW, a social media platform that used to be popular in Hungary, over its full life cycle (2002-2012). This unique dataset allows us to investigate two major geographical features that characterize spatial contagion dynamics: town size described by the urban scaling law \cite{bettencourt2007growth} and distance decay described by the gravity law \cite{liben2005geographic}. 
We find empirical evidence that early adoption is concentrated in large towns and scales super-linearly with town population but late adoption is less concentrated. Diffusion starts across distant big cities such that distance decay of spread is slight and becomes more local over time as adoption reaches small towns in later stages when distance decay becomes strong.

\vskip 0.1in
To better understand the spatial characteristics of complex contagion in social networks, we develop a Bass ABM of new technology's adoption on a sample of the empirical network preserving the community structure and geographical features of connections within and across towns. The data allows us to measure individual adoption thresholds that we can use to parameterize the likelihood of adoption at given fractions of infected social connections. We compare how the ABM and the Bass differential equation (DE) model fit to the empirical urban scaling and distance decay characteristics. Finally, we evaluate model accuracy in predicting the time of local adoption peaks and assess the bias induced by local network structures, or geographical features of towns. These analyses enable us to evaluate the role of geography in complex contagion models at local scales.

\vskip 0.1in
We find that the scaling of the number of earliest adopters with town population is best reflected by the ABM when threshold parameters are incorporated. None of our models can reproduce the high probability of diffusion across distant peers in the early stages of the life-cycle. Certain features of the network within towns - eg. high network density and transitivity - accelerate the ABM diffusion and make predictions of adoption peaks early, which can be overcome when controlling for threshold distributions. Meanwhile, other features of the network - eg. modularity and average path length - delay the prediction of adoption peaks, and cannot be eliminated with the threshold control. Nonetheless, we assess that contagion models cannot cure the bias of physical geography, such as distance from the innovation origin and town size, on the predictions of adoption peaks. 

\vskip 0.1in
The  threshold  mechanisms introduced to the Bass ABM allow us to reproduce aggregated effects in relation to the number of adopters per population size. However, as expected, it is hard to predict the location of the social ties when an adoption occurs. This is in turn, affects the prediction of when the different towns reach their tipping point. Unfolding these aforementioned empirical features, we were able to capture the limitations of the standard model of complex contagion in predicting adoption at local scales and to describe key elements of diffusion in geographical space through the contact of local and distant peers.

\section*{Data}
The social platform analyzed in this work is iWiW, which was a Hungarian online social network (OSN) established in early 2002. The number of users was limited in the first three years, but started to grow rapidly after a system upgrade in 2005 in which new functions were introduced (e.g. picture uploads, public lists of friends, etc.). iWiW was purchased by Hungarian Telecom in 2006 and became the most visited website in the country by the mid-2000s. Facebook entered the country in 2008, and outnumbered iWiW daily visits in 2010, which was followed by an accelerated churn. Finally, the servers of iWiW were closed down in 2014. All in all, more than 3 million users (around 30\% of the country population) created a profile on iWiW over its life-cycle and reported more than 300 million friendship ties on the website. Until 2012, to open a profile, new users needed an invitation from registered members. Our dataset covers the period starting from the very first adopters (June 2002) until the late days of the social network (December 2012). Additionally, it contains home locations of the individuals, their social media ties, invitation ties, and their dates of registration and last login for each of the 3,056,717 users. The last two variables can be used to identify the date of adoption and disadoption (also referred to as churn) on individual level. Spatial diffusion and churn of iWiW have been visualized in Movie S1.

\vskip 0.1in
In previous studies, the data has demonstrated that the gravity law applies to the spatial structure of social ties \cite{lengyel2015geographies}, that adoption rates correlate positively both with town size and with physical proximity to the original location \cite{lengyel2016online}, that users central in the network churn the service after the users who are on the periphery of the network \cite{lorincz2017collapse}, and that the cascade of churn follows a threshold rule \cite{torok2017cascading}. Socio-economic outcomes such as local corruption risk \cite{wachs2019social} and income inequalities \cite{toth2019inequality} have been also investigated with the use of iWiW data. 

\begin{figure*}[!b]
\centering
\includegraphics[width=0.8\linewidth]{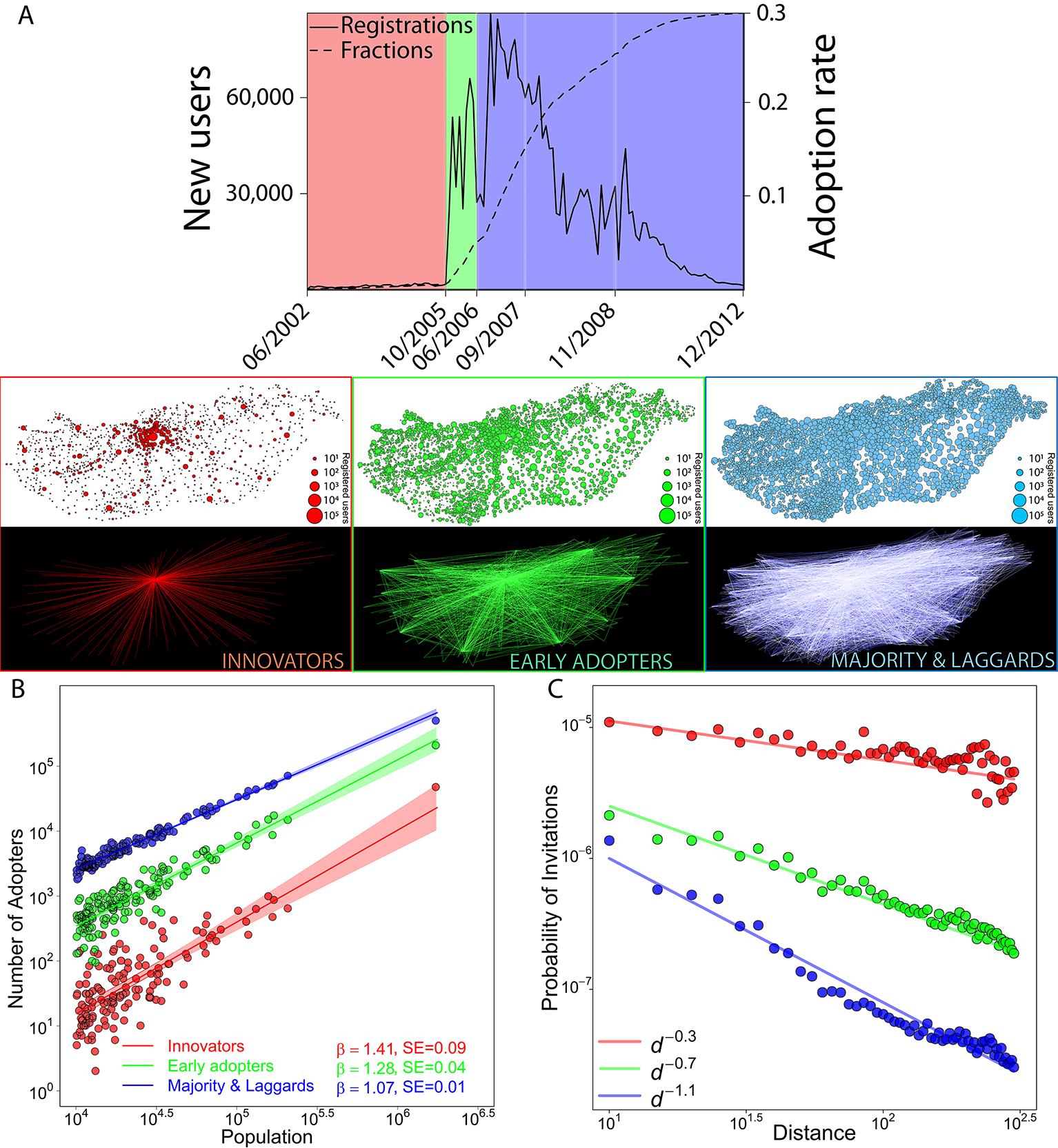}
\caption{\textbf{Spatial diffusion over the OSN life-cycle.} \textbf{A.} \textit{Top:} Number of new users and the cumulative fraction of registered individuals among total population over the OSN life-cycle. Users are categorized by the time of their registration into Rogers's adopter types: (I.) Innovators: first 2.5\%, (II.) Early Adopters: next 13.5\%, (III.) Early and Late Majority and Laggards. \textit{White background maps:} Coloured dots depict towns; their size represent the number of adopters over the corresponding period. \textit{Black background maps:} Links depict the number of invitations sent between towns over the corresponding periods. \textbf{B.} Adoption scales super-linearly with town population. The $\beta$ coefficient of urban scaling denotes very strong concentration of adoption in the Innovator stage and decreases gradually in later stages. Fitted lines explain the variation in Number of Adopters (log) by $R^2=0.63$ (red), $R^2=0.83$ (green), $R^2=0.97$ (blue) \textbf{C.} The Probability of Invitations to distant locations is relatively high in the Inventors stage but decreased over the product life-cycle while diffusion became more local. Exponent fits explain the variation of Probability of Invitations (log) with $R^2=0.24$ (red line), $R^2=0.85$ (green line), $R^2=0.92$ (blue line).}\label{fig:fig1}
\end{figure*}

\section*{Results}
In the first step of the analysis, we empirically investigated the spatial diffusion
over the OSN life-cycle. We categorized the users based on their adoption time for which we applied the rule proposed by Rogers \cite{rogers2010diffusion} that divides adopters as follows: (I.) Innovators: first 2.5\%, (II.) Early adopters: next 13.5\%, (III.) Early Majority: following 34\%, (IV.) Late Majority: next 34\%, and (V.) Laggards: last 16\%. Figure \ref{fig:fig1}A illustrates the number of new users and the cumulative adoption rate (top plot), the spatial distribution of registered users (maps over white background) and the spatial patterns of accepted invitations to register (maps over black background). In the Innovator phase that lasted for three years (in red), adoption occurred in the metropolitan area of Budapest from where the innovation spouted over long distances, reaching the most populated towns first. In the Early Adopters phase (in green) and later in the Majority and Laggards phases (in blue), adoption became spatially distributed and more towns started to spread invitations. 

\vskip 0.1in
The data allow us to demonstrate two major empirical characteristics of spatial diffusion proposed by previous literature \cite{hagerstrand1968innovation}. First, by regressing the number of adopters with town population (both on logarithmic scale) \cite{bettencourt2007growth} we find in Figure \ref{fig:fig1}B that the number of Innovators and Early Adopters ($\beta_{Innovators}=1.41$, CI[1.23;1.59], $\beta_{Early Adopters}=1.28$, CI[1.18;1.37], and $\beta_{Majority\&Laggards}=1.07$, CI[1.04;1.10]) are strongly and significantly concentrated in large towns. Second, in Figure \ref{fig:fig1}C we illustrate the gravity law \cite{liben2005geographic} by stages of the life-cycle by depicting the probability of invitations sent to a new user at distance \textit{d} formulated by ($P^t_d=L^t_d / N^t_a \times N^t_b$), where $L^t_d$ refers to the number of invitations sent at $d$ over stage $t$ while $N^t_a$ and $N^t_b$ denote the number of users who registered in stage $t$ in towns $a$ and $b$ separated by $d$. The strengthening distance decay of invitation links demonstrates that diffusion first bridges distant locations but becomes more and more local over the life-cycle.

\subsection*{Adoption in the Bass diffusion framework}

The Bass diffusion model \cite{bass1969new} enables us to investigate adoption dynamics at global and local scales. This can be done by fitting the cumulative distribution function (CDF) of adoption (shown in Figure \ref{fig:fig1}A) with model CDF. The Bass CDF is defined by $dy(t)/dt = (p+q \times y(t))(1-y(t))$, with $y(t)$ the number of new adopters at time \textit{t} (months), \textit{p} innovation or advertisement parameter of adoption (independent from the number of previous adopters), and \textit{q} imitation parameter (dependent on the number of previous adopters). This nonlinear differential equation can be solved by: 
\begin{equation}
y(t) = m\frac{1-e^{-(p+q)t}}{1+\frac{q}{p}e^{-(p+q)t}}, 
\label{Eq1}
\end{equation}
with $m$ size of adopting population. Eq. \ref{Eq1} described the CDF empirical values with 
residual standard error $RSE=0.0001398$ on $df=125$ and empirical values  \textit{q} = 0.108, CI[0.097;0.12]; \textit{p}= 0.00016, CI[$10^{-4}$;$2\times 10^{-4}$]. We repeated these estimations of the diffusion parameters for every geographic settlement $i$ (called towns henceforth) and consequently estimated $p_i$ and $q_i$. 

\begin{figure*}[!t]
\centering
\includegraphics[width=\linewidth]{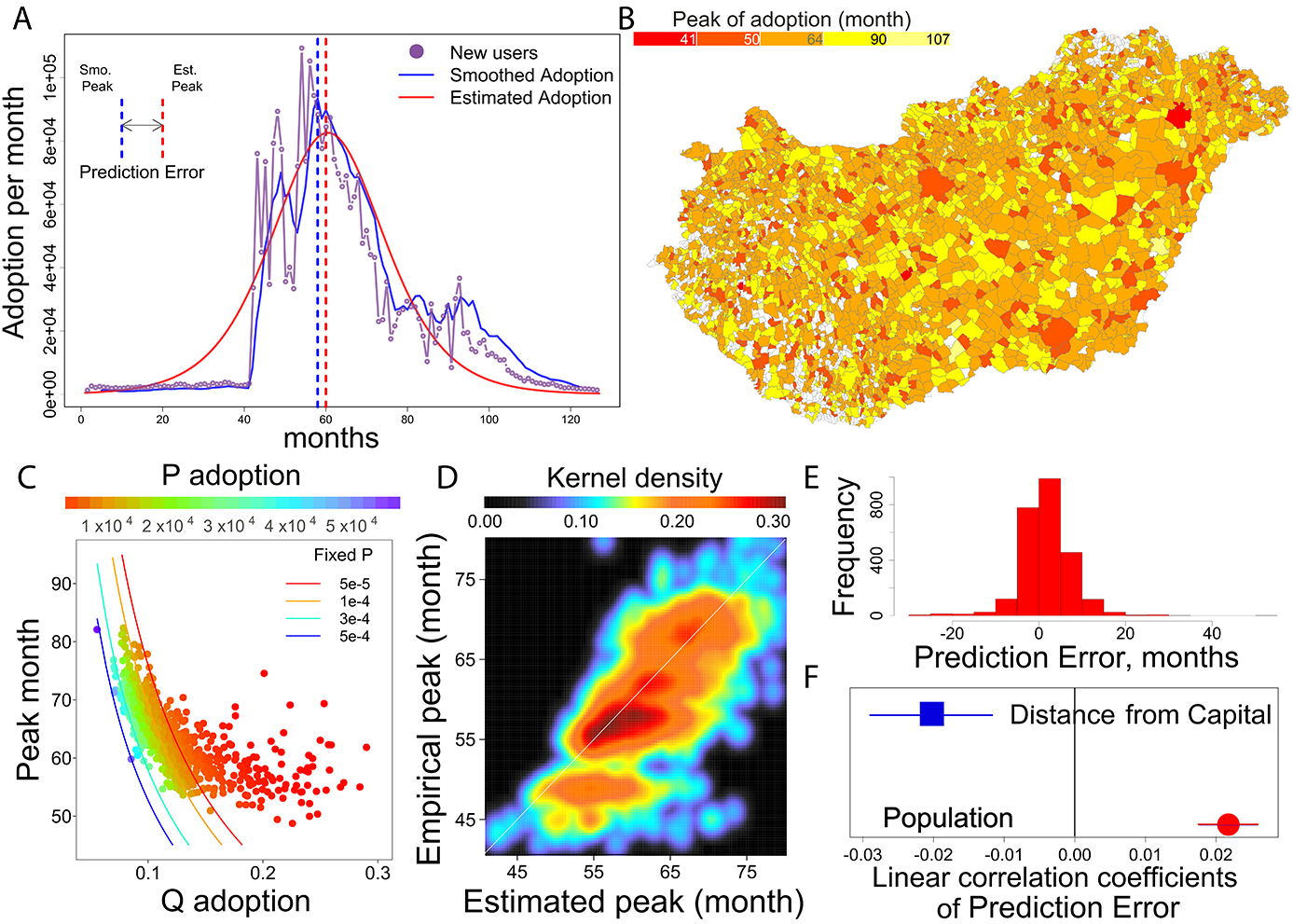}
\caption{\textbf{Adoption peak prediction on local scales with the Bass model.} \textbf{A.} The Bass DE model estimates on the monthly adoption trend and a smoothed empirical adoption trend (3-month moving average) are compared. We investigate the difference between estimated peak month and the peak of the smoothed trend. \textbf{B.} Times of adoption peaks varies across towns.
\textbf{C.} Estimated $p_i$ and $q_i$ result in same adoption peak with fixed $p_i$, except in early adoption cases when $q_i$ is high.
\textbf{D.} Estimated peaks of adoption correlate with empirical peaks of adoption ($p=0.74$). \textbf{E.} Prediction Error in town $i$ is the predicted month of adoption peak by Equation \ref{eq:peak} minus the empirical month of smoothed adoption peak. \textbf{F.} Dots are point estimates of linear univariate regressions and bars depict standard errors. Dependent variable is scaled with its maximum value and independent variables are log-transformed with base 10.}
\label{fig:fig2}
\end{figure*}

\vskip 0.1in
The time of adoption peak \cite{toole2012modeling}, defined by the maximum amount of adoption per month, is an important feature of adoption dynamics. To evaluate the Bass model accuracy on local scales, we investigate Prediction Error, the peak month predicted by the model minus the empirical peak month (smoothed by a 3-month moving average that helps to eliminate noise). Prediction Error is illustrated in Figure \ref{fig:fig2}A. Towns' differences in terms of the time of adoption peak indicate a wide distribution of local deviations from the global diffusion dynamics (Figure \ref{fig:fig2}B), which can be used in statistical analysis. 
The Bass model estimation of the adoption peak $t^\ast_i$ for every town $i$ is:
\begin{equation}
t^\ast_i=\frac{\ln p_i + \ln q_i}{p_i+q_i}
\label{eq:peak}
\end{equation}
and is positively correlated with the empirical peaks in Figure \ref{fig:fig2}D ($\rho=0.742$, CI[0.725;0.759]).

\vskip 0.1in
In case we keep one of $p_i$ and $q_i$ parameters fixed, adoption becomes faster as the other increases (Figure \ref{fig:fig2}C). Furthermore, towns diverge from Eq.~\ref{eq:peak} for peak times in months 50-60 (Figure \ref{fig:fig2}D), corresponding to low $p_i$ and large $q_i$ (Figure \ref{fig:fig2}C). This suggests that the innovation term in the Bass model is lower and the process is driven by imitation in towns where diffusion happens at the primitive stage. On average, peaks in towns predicted by Eq. \ref{eq:peak} are 1.76 months later, with a 95\% confidence interval [1.54; 1.98], than empirical peaks (Figure \ref{fig:fig2}E). Prediction is late in large towns but is early in towns distant from Budapest that are also smaller than average (correlation between population and distance is $\rho=-0.32$) CI[-0.35;-0.28] (Figure \ref{fig:fig2}F). Population correlates with both Eq.~\ref{eq:peak} parameters (with $p_i$, $\rho=0.11$ CI[0.07;0.14] and with $q_i$, $\rho=-0.34$ CI[-0.34;-0.30]). The correlation between Bass parameters, peak prediction and town characteristics are reported in Supporting Information 2. 
\vskip 0.1in
Although parameters are estimated for every town separately, physical geography still influences model prediction. 
An important limitation of modelling local adoption with Bass DE is that towns are handled as isolates. To disentangle the role of geography in diffusion, we need models that can consider connections between locations.

\subsection*{A complex diffusion model}

We further investigated the spreading of adoption on a social network embedded in geographical space connecting towns and also individuals within these towns via the ABM version of the Bass model. We used the social network observed in the data by keeping the network topology fixed at the last timestamp without removing the churners, using this as a proxy for the underlying social network. This approximation is a common procedure to model diffusion in online social networks when the underlying social network cannot be detected \cite{karsai2016local}. The ABM is tested on a $10\%$ random sample of the original data (300K users) by keeping spatial distribution and the network structure stratified by towns and network communities. The latter were detected from the global network using the Louvain method \cite{blondel2008fast}. We show in Supporting Information 4 that samples of different sizes have almost identical network characteristics and these are very similar to the full network as well.

\begin{figure*}[!t]
\centering
\includegraphics[width=\linewidth]{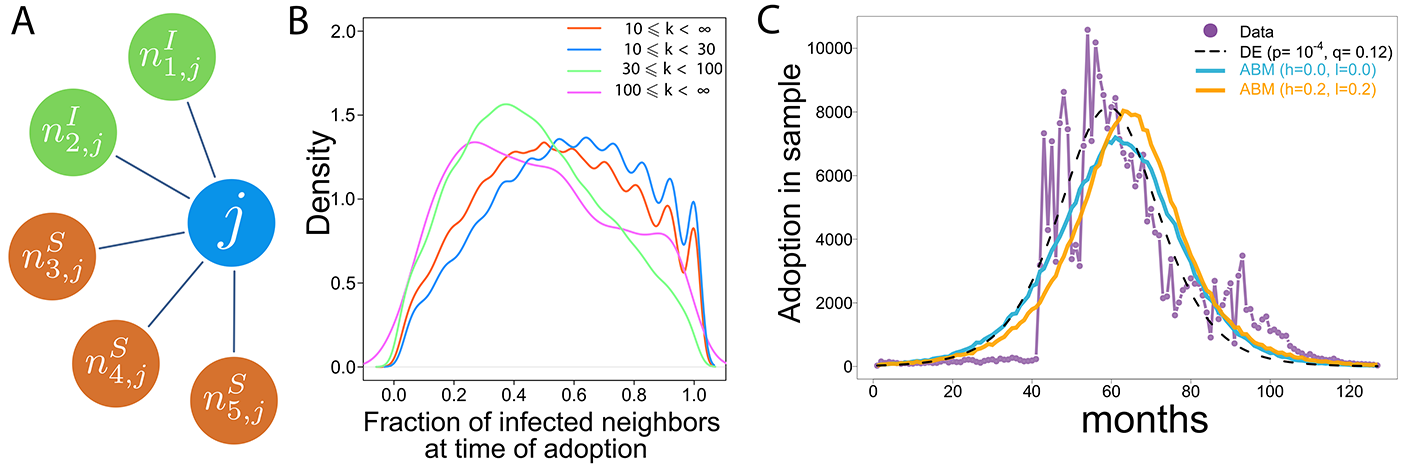}
\caption{\textbf{Model of complex contagion.} \textbf{A.} Network topology and peers influence in the Bass ABM. A sample individual $j$ has two infected neighbors $n_j^I$ who have already adopted the innovation and three susceptible neighbors who have not adopted yet $n_j^S$. \textbf{B.} The distribution of adoption thresholds. Fraction of infected neighbors at time of adoption illustrate that most individuals adopt when half of their neighbors have already adopted. This fraction is smaller for high degree (k>30) individuals. \textbf{C.} ABM adoption curves assuming linear ($h=0.0, l=0.0$ in blue) and non-linear ($h=0.2, l=0.2$ in red) functions of infected neighbor ratio predict slower adoption than Bass DE.
}
\label{fig:fig3}
\end{figure*}

\vskip 0.1in
In the ABM, each agent $j$ has a set of neighbors $n_j$ taken from the network structure (Figure \ref{fig:fig3}A) and is characterized by a status $F_j(t)$ that can be susceptible for adoption $S$ or infected $I$ (already adopted). Once an agent reaches the status $I$, it cannot switch back to $S$. To reflect reality, the users that adopted in the first month in the real data were set as infected $I$ in $t=1$. The process of adoption $F_j(t)=S \rightarrow F_j(t+1)=I$ is defined as:
\begin{equation}
F_j(t+1)=\begin{cases}
I & \mbox{if } U(0,1)_{jt}< \hat{p}^{\mbox{ABM}}+T(\mathcal{N}_j(t),h,l)\times\mathcal{N}_j(t)\times \hat{q}^{\mbox{ABM}} \\
S &\mbox{otherwise}
\end{cases}
\label{eq:abm}
\end{equation}

\noindent where $U(0,1)_{jt}$ is a random number picked from a uniform distribution for every agent $j$ in each $t$. $\hat{p}^{\mbox{ABM}}$ denotes adoption probability exogenous to the network and $\hat{q}^{\mbox{ABM}}$ is adoption probability endogenous to the network. In order to focus on the role of network structure in diffusion, $\hat{p}^{\mbox{ABM}}$ and $\hat{q}^{\mbox{ABM}}$ are kept homogeneous for all $j$ in the network. Consequently, the process is driven by the neighborhood effect $\mathcal{N}_j(t)$ defined as:
\begin{equation}
\mathcal{N}_j(t)=\frac{\#n^I_j(t)}{\#n^I_j(t)+\#n^S_j(t)} 
\end{equation}
where $\#n^I_j(t)$ is the number of infected neighbors and $\#n^S_j(t)$ is the number of susceptible neighbors at $t$.

\vskip 0.1in
The distribution of $\mathcal{N}_j(t)$ at the time of adoption carries information about adoption dynamics in the social network \cite{karsai2016local}. Figure \ref{fig:fig3}B suggests that the probability of adoption in our case is the highest when $\mathcal{N}_j(t)$ is around 0.5 (in case 10 $\leq k \leq \infty$) and decreases when $\mathcal{N}_j(t)$ is close to 0 or 1. To reflect on this empirical finding in the ABM, we introduce the transformation function $T(\mathcal{N}_j(t),h,l)$ on $\mathcal{N}_j(t)$ defined by
\begin{equation}
    T(\mathcal{N}_j(t),h,l) = -4(h+l)\cdot \mathcal{N}_j(t)^2+4(h+l)\cdot \mathcal{N}_j(t) +1-l.
\label{eq:t}
\end{equation}
where $h$ controls the relative importance of $\mathcal{N}_j(t)=0.5$ and $l$ controls the decrease of the adoption probability at $\mathcal{N}_j(t)=0$ and $\mathcal{N}_j(t)=1$. Both of parameters $h$ and $l$ are considered in order to find optimum model descriptions of spatial adoption.

\vskip 0.1in
This definition of the process implies that users are assumed to be identically influenced by advertisements and other external factors and are equally sensitive to the influence from their social ties that are captured by the fraction of infected neighbors $\mathcal{N}_j(t)$. The decision regarding adoption of innovation or postponing this action is an individual choice that is assumed to be random. This model belongs to the complex contagion class \cite{centola2007complex,centola2010spread} because adoption over time is controlled by the fraction of infected neighbors \cite{watts2002simple,karsai2016local}. As the fraction of infected neighbors increases, the agent becomes more likely to adopt the innovation. Supporting Information 4 describes the calibration of $\hat{p}^{\mbox{ABM}}$ and $\hat{q}^{\mbox{ABM}}$, and explain how $h$ and $l$ parameters were selected.

\vskip 0.1in
We set Bass parameters in the ABM to their calibrated values $\hat{p}^{\mbox{ABM}}=0.0002$ and $\hat{q}^{\mbox{ABM}}=0.12$ that are close to the estimated values using Eq.\ref{Eq1} on the ABM sample (reported in Figure \ref{fig:fig3}B) as suggested by \cite{xiao2017bass}. Two ABMs are considered. ABM ($h=0.0$, $l=0.0$) assumes that adoption probability increases linearly with $\mathcal{N}_j(t)$. ABM ($h=0.2$, $l=0.2$) assumes a non-linear influence of $\mathcal{N}_j(t)$ on adoption probability. Supporting Information 4 illustrates $T(\mathcal{N}_j(t),h,l)$ with parameters h=0.2 and l=0.2, and it's relation with the empirical threshold distribution and explains how parameters ($h=0.2$, $l=0.2$) change adoption probability in the ABM compared to the case when $h=0.0$ and $l=0.0$. 

\vskip 0.1in
In Figure \ref{fig:fig3}C, we report global adoption trends after running both ABM 10 times and calculating average values of these realizations over time-steps $t=(1,120)$ that reflects the months taken from the real data. Both ABM($h=0.0$, $l=0.0$) (solid blue line) and ABM($h=0.2$, $l=0.2$) (solid orange line) are faster in the early phase (before month 40) than in reality, which is due to the extraordinary tipping point around month 40 that is difficult to fit. ABM($h=0.2$, $l=0.2$) is closer to reality in this early phase while ABM($h=0.0$, $l=0.0$) follows the DE trend until month 40. Comparing to ABM($h=0.0$, $l=0.0$), ABM($h=0.2$, $l=0.2$) is faster from month 40, has an adoption volume at its peak comparable to the DE estimate, and decline faster after it's peak. The peak predicted by DE is at month 59, by ABM($h=0.0$, $l=0.0$) is at month 61, and by ABM($h=0.2$, $l=0.2$) is at month 63; whereas the empirical peak smoothed with 3 months moving average is at month 58. Adoption in ABM($h=0.0$, $l=0.0$) fit o adoption in DE with $\chi^2=15,621, p=4 \time 10^{-4}$ while ABM($h=0.2$, $l=0.2$) fit to adoption in DE with $\chi^2=15,748, p=4 \time 10^{-4}$. These initial comparisons suggest that ABM($h=0.0$, $l=0.0$) can capture early adoption dynamics better than ABM($h=0.0$, $l=0.0$), while the peak of adoption might be better reproduced by ABM($h=0.0$, $l=0.0$). 

\subsection*{Local adoption in the ABM}

To better understand the differences between DE and ABM versions, we move now from the global trend to local scales and compare DE that is informed by location-specific $p_i$ and $q_i$ but cannot incorporate networks with ABM that can control networks but has homogenous $\hat{p}^{\mbox{ABM}}$ and $\hat{q}^{\mbox{ABM}}$. The introduction of $T(\mathcal{N}_j(t),h,l)$ enables us to investigate how controlling for the threshold distribution improves ABM predictions at local scales compared to data and the DE estimations.

\begin{figure*}[!b]
\centering
\includegraphics[width=\linewidth]{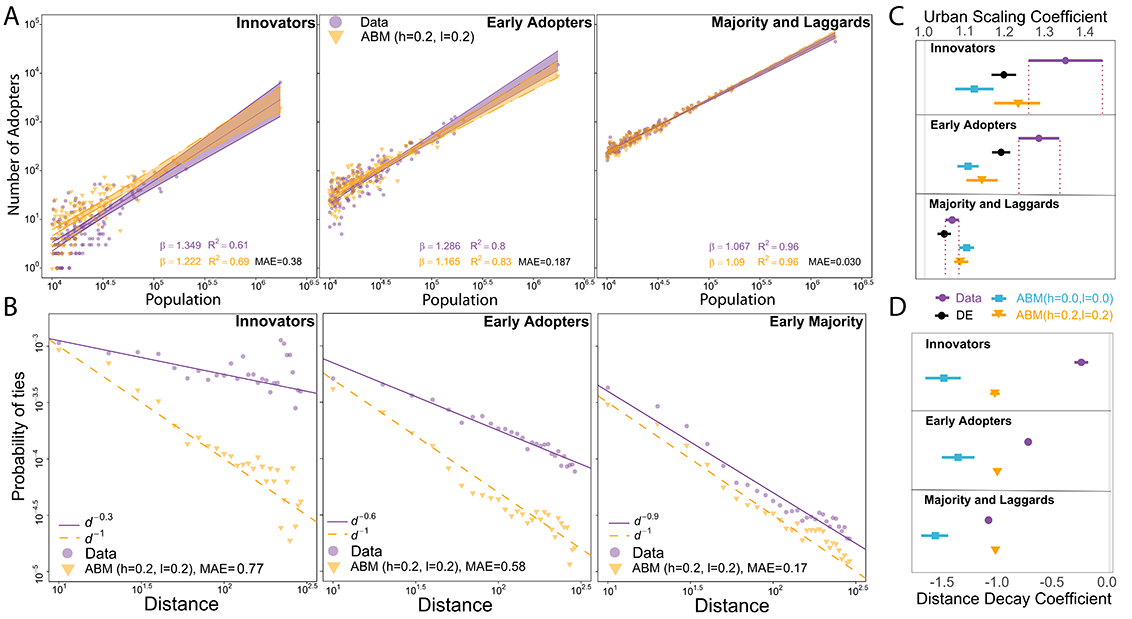}
\caption{\textbf{Urban scaling and distance decay in the ABM.} \textbf{A.} Urban scaling of adoption in the ABM(h=0.2, l=0.2) and in the empirical data across the product life-cycle. Solid lines denote linear regression estimation and shaded areas are 95\% confidence intervals. The ABM(h=0.2,l=0.2) significantly over-predicts the number of Innovators in small towns. Urban scaling $\beta$ in the Early Adopters phase is still smaller in the ABM(h=0.2,l=0.2)  than in the empirical data. \textbf{B.} The distance decay of social ties of Innovators and Early Adopters is larger in the ABM(h=0.2, l=0.2) than in reality and only becomes similar in the Early Majority phase. \textbf{C.} Empirical urban scaling coefficients are declining over the life-cycle that is best captured by the ABM(h=0.2, l=0.2) prediction. Markers denote point estimates and horizontal lines denote standard errors. \textbf{D.} Empirical distance decay coefficients are declining over the life-cycle that are not captured by the models. Markers denote point estimates and horizontal lines denote standard errors.} 
\label{fig:fig4}
\end{figure*}

\vskip 0.1in
A major challenge in spatial diffusion modeling is the unknown spatial distribution of Innovators and Early Adopters that need to be predicted by the model; however, as a paradox, this spatial distribution is a prerequisite of accurate prediction of local adoption peaks in social networks\cite{toole2012modeling}. 

To overcome this limitation, we empirically analyze how the ABM captures spatial distribution of adoption in three phases of product life-cycle. In Figures \ref{fig:fig4}A and C we compare how the number of adopters observed in the data and predicted by the model scale with the town population \cite{bettencourt2007growth, deville2016scaling} by using the $\beta$ coefficient of the linear regression in towns with more than $10^4$ inhabitants. Because both ABM(h=0.0, l=0.0) and ABM(h=0.2, l=0.2) are faster than real adoption in the first 40 months but are slower than DE and following Rogers\cite{rogers2010diffusion} we define Innovators and Early Adopters as the first 2.5\% and the next 13.5\% of adopters. This enables us to compare spatial distribution of Innovators and Early Adopters between the ABMs, Bass DE and reality regardless of temporal differences in the global trend. 

\vskip 0.1in
An empirical superlinear scaling measured in the sampled Data in the Innovator and Early Adopter phases indicates strong urban concentration of diffusion during the early phases of adoption, already reported in Figure \ref{fig:fig1} on the full network. Supporting Information 5 demonstrates that the urban scaling estimation is robust against introducing various indicators of town development or demographics. To compare Bass ABM and Bass DE approaches, we re-estimate Eq.\ref{Eq1} for every town in the sample and estimate monthly adoption that can enter the scaling regression. Figure \ref{fig:fig4}C reveals that ABM(h=0.2, l=0.2) follows the changes in empirical urban scaling somewhat better both in terms of $\beta$ and fit to empirical adoption than ABM(h=0.0, l=0.0) that has an urban scaling $\beta$ of adoption around 1.1 in all phases of the life-cycle. The scaling coefficient of ABM(h=0.2, l=0.2) is within the margin of error in the Innovator and Majority and Laggards phases; in the former this is due to the large standard error of empirical scaling coefficient. ABM(h=0.2, l=0.2) partly outperforms the DE estimation that only captures scaling of Early Adopters better. However, we find in Figure \ref{fig:fig4}A that in the Innovator phase of the life-cycle, the ABM predicts more adoption in small towns and less in large towns compared to reality and predicts smaller adoption volumes in large towns in the Early Adopters stage. What happens is that the ABM interchanges individuals' early adoption in large towns with early adoption in small towns such that much more small town users get into the first 2.5\% than in reality. This is a bit less striking when adoption probability is increased at most frequent individual thresholds in ABM(h=0.2, l=0.2), which probably slows ABM adoption down in small towns. Confidence intervals of urban scaling coefficients plotted in Figure \ref{fig:fig4}C can be found in Supporting Information 6.

\vskip 0.1in
Turning to the role of distance in diffusion over the life-cycle, Figure \ref{fig:fig4}B compares the distance of influential peers, measured as the probability that Innovators, Early Adopters, and Early Majority \cite{rogers2010diffusion} have social connections at distance $d$ \cite{liben2005geographic, lengyel2015geographies, scellato2010distance, onnela2011geographic, wang2009understanding} in the ABM(h=0.2,l=0.2) versus in the empirical data. 

Ties of Innovators have a very week distance decay, which intensifies for Early Adopters and even more for Early Majority. The intensifying role of distance measured here resembles distance decay measurement by invitation data (see Figure \ref{fig:fig1}) and confirms that innovation spreads with high propensity to distant locations during the early phases of the life-cycle \cite{hagerstrand1968innovation}. However, neither ABM(h=0.2,l=0.2) nor ABM(h=0.0,l=0.0) are able to handle the changing role of distance. Instead, distance decay in both ABMs are rather stable across these three phases of the life-cycle (Figure \ref{fig:fig4}C). Unfortunately, we are not able to compare these patterns to DE estimations, since the distance decay of social connections can't be inferred on with the DE method due to the lack of individual predictions. Our findings imply that ABM replaces distant contagion with proximate contagion in the early phases of the life-cycle. Innovators are mostly found in distant large towns. Even though they are connected to each other, these connections might be bridges across communities that slows complex contagion in the ABM \cite{centola2007complex}. 

\begin{figure*}[!b]
\centering
\includegraphics[width=\linewidth]{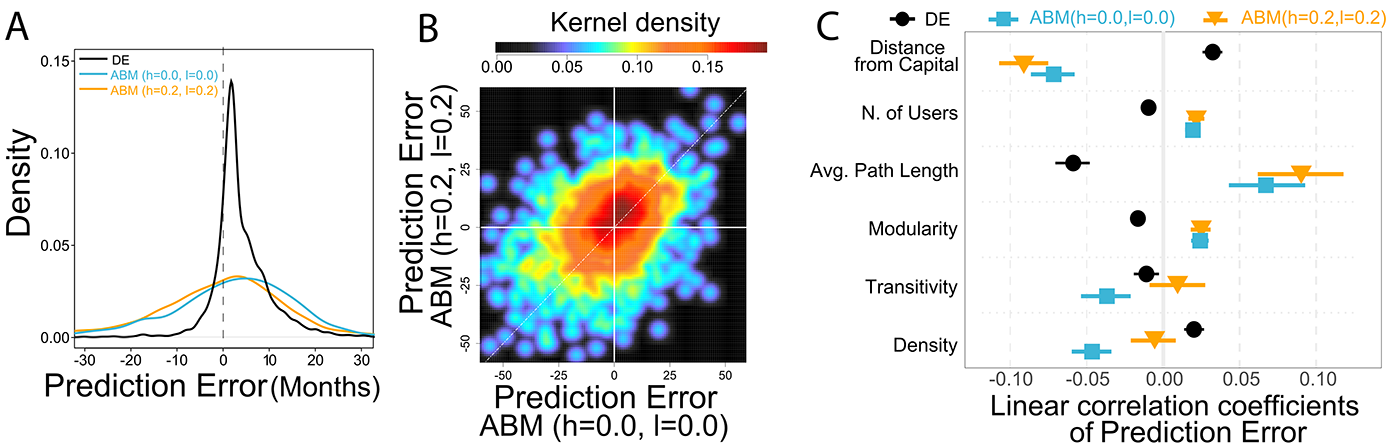}
\caption{ABM in predicting adoption peaks in towns. \textbf{A.}  Prediction Error in town $i$ is the peak month predicted by the ABM minus the month of empirical peak. Negative Prediction Error denotes early prediction and positive means late prediction. \textbf{B.} Correlation of Prediction Error of ABM(h=0.0, l=0.0) and ABM(h=0.2, l=0.2) ($\rho=0.39$, CI[0.36;0.43]). \textbf{C.} Estimation of town-level Prediction Error of ABM versions via a simple linear regression. Independent variables are characteristics from networks within towns and geographical characteristics of towns. Symbols represent point estimates and horizontal lines denote standard errors.}
\label{fig:fig5}
\end{figure*}

\vskip 0.1in
Adoption peaks typically happen in the Early- and Late Majority phases of the life-cycle, for which ABM(h=0.2, l=0.2) adoption predicts the aggregated number of adopters in towns well (Figure\ref{fig:fig4}A). To understand how accurate the peak time predictions are, we analyze determinants of ABM Prediction Error as already done in Figure\ref{fig:fig2}E for the Bass model on the full network. In case of ABM(h=0.0, l=0.0), the predicted month of adoption peak matches the observed month of adoption peak in the data with 95\% confidence interval [-1.69; -0.46]; indicating that the ABM(h=0.0, l=0.0) predicts adoption peaks early in most towns (Figure \ref{fig:fig5}A). However, peaks predicted by ABM(h=0.2, l=0.2) are 1.74 months late on average with 95\% confidence interval [1.16; 2.32]. Peaks predicted by the Bass DE are 3.89 months late on average with 95\% confidence interval [3.62; 4.16]. Prediction error values of the ABMs are correlated (Figure \ref{fig:fig5}B). However, there are towns, where prediction is early in ABM(h=0.0, l=0.0) and is late in ABM(h=0.2, l=0.2) and vice versa.

\vskip 0.1in
In order to analyze the role of network structure in local adoption dynamics in the Majority phase, we correlated the town-level Prediction Errors with several town-level network properties (Figure \ref{fig:fig5}C). Density, the fraction of observed connections among all possible connections in the town's social network; and Transitivity, the fraction of observed triangles among all possible triangles in the town's social network, are claimed to facilitate diffusion \cite{centola2007complex}. On the other hand, complex contagion is more difficult in networks with modular structure, when social links between network communities are sparse, and in networks with long paths, when the distance of nodes within the town's social network is large. In fact, the ABM(h=0.0, l=0.0) predicted the peak of adoption early in the towns where Density and Transitivity are relatively high (Figure \ref{fig:fig5}C). Influencing the probability of adoption according to the adoption threshold distribution in ABM(h=0.2, l=0.2), however, cures this bias as the co-efficients of Density and Transitivity become non-significant. ABM modification does not cure the delaying influence of Modularity and Average Path Length. These latter co-efficients of ABM(h=0.0, l=0.0) and ABM(h=0.2, l=0.2) are within estimation error. We also find that Assortativity, the index of similarity of peers in terms of adoption time \cite{newman2003mixing} delays adoption of large towns, which we discuss in detail in Supporting Information 7. DE Prediction Error estimations are illustrated for the reasons of comparison. We find that local network estimations on DE Prediction Error are not corresponding with ABM estimations and are even counter-intuitive from a network diffusion perspective. These are in line with expectations because DE prediction is not allowed to use information on the local network structure. This finding support our claim that network-based models are needed to better understand diffusion on networks. Confidence intervals of coefficients plotted in Figure \ref{fig:fig5}C can be found in Supporting Information 8.

\vskip 0.1in
Finally, we observe that geographical characteristics, Population (measured here by number of users in the ABM sample) and Distance (measured by Euclidean distance from Budapest) influence the accuracy of ABM peak prediction. Like we found in the case of the Bass DE model on the full network in Figure \ref{fig:fig2}F, prediction is late in large towns but is early in towns distant from Budapest, that are significantly smaller in terms of population than average (see multiple regression results in Supporting Information 9). Point estimates of ABM(h=0.0, l=0.0) and ABM(h=0.2, l=0.2) are not significantly different from each other but are significantly different from DE estimates on the sample. These latter estimations are reported only for the sake of comparison. The DE coefficients seem to be biased by the sampling process, and thus the difference between coefficients in Figures \ref{fig:fig2}F and \ref{fig:fig5}C, and are not robust against regressing them together in a multiple regression framework (see Supporting Information 9). The ABM coefficients confirm that geography has a role in the complex diffusion of innovations. We suggest social contagion models to incorporating town size and geographical distance between peers in order to improve accuracy of local adoption prediction.

\section*{Discussion}

Taken together, we studied spatial diffusion over the life-cycle of an online product on a country-wide scale. 
By combining complex diffusion with empirical threshold distribution, we proposed a stochastic modeling framework that allows for spontaneous adoption in the network and is able to explore how geography influences model accuracy in capturing local adoption trends. The model does not perfectly predict how adoption rates scale with a city’s population, especially in the early stages of the life cycle. This is to some extent due to the fact that the standard model assumes a linear relation between adoption probability and the share of neighbors already active on the OSN. In reality, the relation between individual adoption probability and adoption rates of neighbors is nonlinear: we observe that adoption rates accelerate for intermediate, but decelerate at very high adoption levels by neighbors. Once the ABM takes this into consideration, it's fit to the observed urban scaling of adoption in the early life cycle periods improves. This step eliminates the influence of dense and transitive local networks as well that would otherwise accelerate adoption peaks in towns too early. 

\vskip 0.1in
One of our most important empirical findings is the changing distance decay of diffusion.  
In fact, 
contagion in the early stages of the product life-cycle occurs mostly between distant locations with larger populations. This new aspect could not be captured by the model, indicating that it needs theoretical extension. The superlinear relation of Innovators and Early Adopters as a function of the town population highlights the importance of urban settlements in the adoption of innovations 
that corresponds with the early notion of Haegerstrand \cite{hagerstrand1968innovation}. Adoption peaks initially in large towns and then diffuses to smaller settlements in geographical proximity. We find that town population and distance from the original location of innovation bias predictions of adoption peak in all models. These findings call for incorporating geography into future models of complex contagion.

\vskip 0.1in
Unlike many of the previous work on social networking cites that investigate a large selection of OSNs \cite{ribeiro2014modeling} or a dominant OSN entering many countries \cite{kassa2018large}, our results are limited to a specific product in a single country. In this regard, future research shall investigate how various types of online products diffuse across space and social networks and in different countries. For example, complex products, which has been reported to scale super-linearly with city size \cite{gomez2016explaining, balland2020complex} might diffuse across locations differently than non-complex products due to the difficulties to adopt complex technologies and knowledge. Technologies compete with each other, which is completely missing from our understanding on spatial diffusion in social networks. Some of the technologies dominate over long periods but when quitting becomes collective, their life-cycle ends \cite{ribeiro2014modeling,kairam2012life,kloumann2015lifecycles}. Recent studies have shown that both adopting  and quitting the technology follow similar diffusion mechanisms \cite{garcia2017understanding, torok2017cascading}. However, the geography of how churning is induced by social networks is still unknown. 

\vskip 0.1in
Future work on spatial diffusion of innovation in social networks has to tackle the difficulty of modeling individual adoption behavior embedded in geographical space. One of the challenges is that individuals are heterogenous regarding adoption thresholds that is non-trivially related to the formation and spatial structure of social networks. Individuals who are neighbors in the social network are likely to be located in physical proximity as well, but this is not always the case\cite{liben2005geographic}. Further, network neighbors typically are alike in terms of adoption thresholds \cite{aral2009distinguishing}. Thus, it is not clear whether social influence has a geographical dimension or we can think of it using a space-less network approach. We propose that investigating and incorporating the distance decay in social influence modeling might help us understanding spatial diffusion of innovation better.

\section*{Methods}

Nonlinear least-square regression with the Gauss-Newton algorithm was applied to estimate the parameters in Eq. \ref{Eq1}. In order to identify the bounds of parameters search, this method needs starting points to be determined, which were \textit{p\textsubscript{i}} = 0.007 and \textit{q\textsubscript{i}} = 0.09 for Eq. \ref{Eq1}. 

\vskip 0.1in
Identical estimations were applied in a loop of towns, in which the Levenberg-Marquardt algorithm \cite{more1978levenberg} was used with maximum 500 iterations. This estimation method was applied because the parameter values differ across towns, and therefore town-level solutions may be very far from the starting values set for the country-scale estimation. Initial values were set to $p_i$ = 7*10\textsuperscript{-5} and $q_i$ = 0.1 in Eq. \ref{Eq1}.

\vskip 0.1in
To characterize urban scaling of adoption and churn in Figures \ref{fig:fig1} and \ref{fig:fig4}, we applied the ordinary least squares method to estimate the formula $y(t) = \alpha + \beta x $, where \textit{y(t)} denotes the logarithm (base 10) of accumulated number of adopters over time period \textit{t}, and \textit{x} is the logarithm (base 10) of the population in the town. R-squared values have been applied to the  variance of the log-transformed dependent variable.

\section*{Data availability}
Data tenure was controlled by a non-disclosure agreement between the data owner and the research group. The access for the same can be requested by email to the corresponding author.

\section*{Code availability}
ABM simulation and parameter calibration codes have been written in Python and have been reposited at \url{https://github.com/bokae/spatial_diffusion}. All other codes to produce the results have been written in R. These latter codes are available upon request at the corresponding author and will be reposited before publication.

\section*{Acknowledgements}

Balazs Lengyel acknowledges financial support from the Rosztoczy Foundation, the Eötvös Fellowship of the Hungarian State, and from the National Research, Development and Innovation Office (KH 130502). Riccardo Di Clemente as Newton International Fellow of the Royal Society acknowledges the support of The Royal Society, The British Academy, and the Academy of Medical Sciences (Newton International Fellowship, NF170505). János Kertész acknowledges funding received from the SoBigData++ H2020 grant (ID: 871042) and from the Hungarian Scientific Research Fund (OTKA K-129124).

\section*{Author contributions statement}

B.L. and M.G. designed the research, B.L., E.B. and R.D.C. conceived the experiments, B.L., E.B., R.D.C., J.K and M.G. analyzed the results. All authors wrote and reviewed the manuscript. 

\section*{Additional information}

\textbf{Competing interests} The authors declare no competing interests. 

\newpage

\renewcommand{\thefigure}{S\arabic{figure}}
\renewcommand{\thetable}{S\arabic{table}} 
\renewcommand{\theequation}{S\arabic{equation}} 
\setcounter{figure}{0}
\setcounter{table}{0}
\setcounter{equation}{0}

\section*{Supporting Information 1: Spatial diffusion and churn over the product life-cycle}

Video on spatial diffusion and churn of iWiW. Nodes denote towns and links represent invitations sent across towns between 2002 and 2012 on a monthly basis. The size of nodes illustrates the number of users who registered in the town by the given month and the color depicts the share of those registered users who still logged in. Adoption started in Budapest (the capital) and was followed first in its surroundings and other major regional subcenters. The vast majority of invitations have been sent from Budapest in the initial phase of diffusion and subcenters started to transmit spreading when diffusion speeded up in the middle of the life-cycle. A decisive fraction of users logged in to the website even after Facebook entered the country in 2008. Collective churn started in 2010 and the rate of active users dropped quickly in most of the towns. Exceptions are small villages in the countryside, where people have difficulties to adopt new waves of social media innovation. 

For the video on spatial diffusion and churn, go to \url{https://vimeo.com/251494015}

\newpage

\section*{Supporting Information 2: Correlation of Bass model predictions and geographical characteristics}

Prediction Error correlates negatively with peak of adoption indicating that Bass prediction of peaks works better in towns that adopt late. Town size and Distance from the Capital are negatively correlated with each other ($\rho=-0.32$). We find that $q_i$ is significantly smaller in large towns than in small towns. There is a significant negative correlation between the month of Predicted Peak and $q_i$; while this correlation with $p_i$ is positive. Standard errors of $p_i$ and $q_i$ correlate strongly with the respective parameters. Further correlations of $SE q_i$ indicate that estimation of $q_i$ is significantly more accurate in towns where adoption peaks late but is less accurate in towns that are far from Budapest.

\begin{figure}[!htb]
\centering
\includegraphics[width=0.9\linewidth]{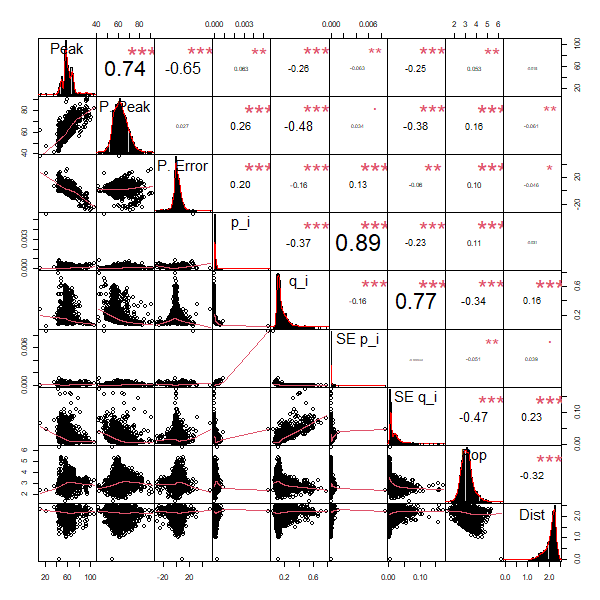}
\caption{Pearson correlation coefficients of Bass model and geographical characteristics of towns. Peak denotes month of observed adoption peak in towns; P. Peak is the predicted month of adoption peak by the Bass model; P. Error is predicted month of adoption peak minus the empirical peak; $p_i$ and $q_i$ denote Bass model parameters; $SE p_i$ and $SE q_i$ are standard errors of the estimated parameters; Pop denotes $log_{10}$ of town population and Dist denotes $log_{10}$ kilometers from Budapest.}
\label{fig:SI2}
\end{figure}

\newpage

\section*{Supporting Information 3: Network sampling for ABM}

To model diffusion in the empirical social network, we sample the full network of 3 Million nodes by keeping the distribution of nodes according to locations and network communities. This is done by identifying the community structure of the full network with the Louvain algorithm and assigning every node into one community. Then, we take 5\%, 10\%, 20\% samples and stop sampling when the p-value of the Kolmogorov-Smirnov test comparing both town and community distributions of the sampled and full node lists is larger than 0.95. Finally, we connect the nodes with ties that link them in the full network and exclude those nodes that are not part of the giant component.

\vskip 0.1in
In Table \ref{tab:sample}, we compare structural characteristics of the 5\%, 10\%, and 20\% sample networks with the full network. Density of links in the sampled networks are on the same magnitude as the full network. However, the smaller sample we take the higher density. Global clustering (the ratio of closed triangles among all possible triangles) is identical across samples, which is around half of the full network. The fraction of links that connect individuals across towns are identical in the samples and the full network.

\begin{table}[h] \centering 
  \caption{Characteristics of sampled networks} 
  \label{} 
\begin{tabular}{|l|l|l|l|l|}
\hline
\textbf{Sample}          & \textbf{5\%} & \textbf{10\%} & \textbf{20\%} & \textbf{100\%} \\ \hline
Nodes in Giant Component & 128,590      & 271,941       & 564,134       & 3,050,988      \\ \hline
Links                    & 675,227      & 2,712,588     & 10,799,507    & 279,708,125    \\ \hline
Density                  & $8.167 \times 10^{-5}$ & $7.33 \times 10^{-5}$   & $6.78 \times 10^{-5}$   & $6.01 \times 10^{-5}$    \\ \hline
Global clustering               & 0.09         & 0.09          & 0.09          & 0.17           \\ \hline
Links across towns, \%   & 50.1\%       & 51.2\%        & 51.1\%        & 51.1\%         \\ \hline
\end{tabular}
\label{tab:sample}
\end{table}

\begin{figure}[!htb]
\centering
\includegraphics[width=\linewidth]{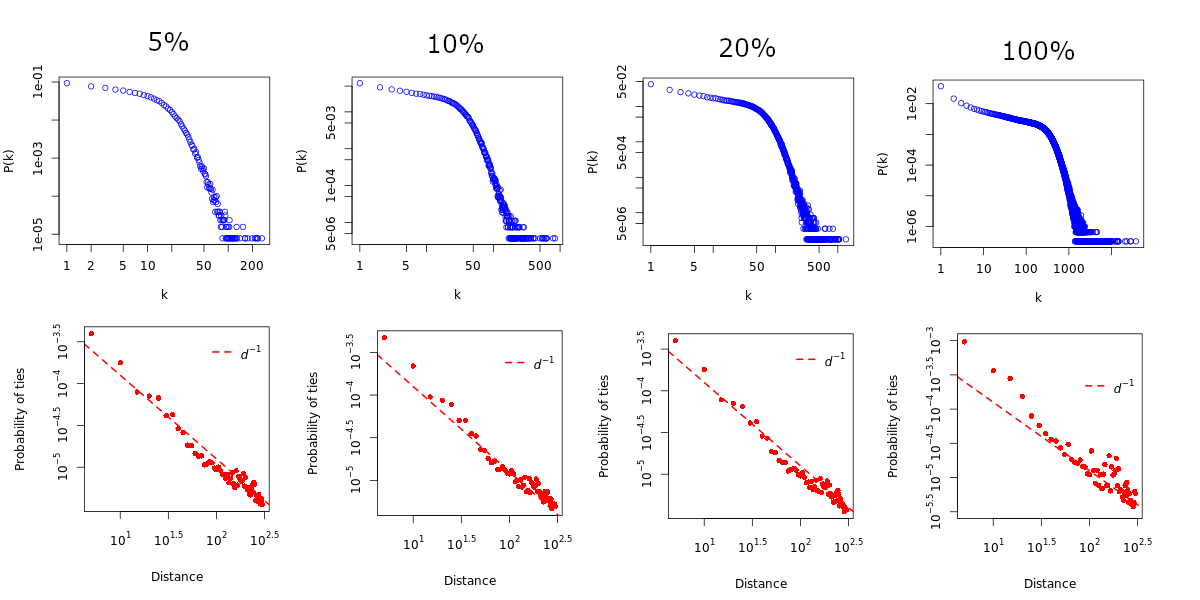}
\caption{Degree distribution and distance decay in the sampled networks}
\label{fig:SI3}
\end{figure}

In Figure \ref{fig:SI3}, we plot degree distribution and distance decay of connections for each sample and the full network. The sample degree distributions lack the high probability of low-degrees (k<10) that is an interesting characteristic of the full network. Further, the probability of ties at short distances ($d<10^{1.5}$) deviate positively from the generally observed distance decay in the full network. This deviation is present in the sample networks as well, but only to a lesser extent. 

In sum, by taking the 10\% sample of the full network, we cannot fully represent the fraction of low degree nodes and short-distance linkages. Consequently, Density is higher and Global Clustering is lower in the sampled network than in the full network. In our understanding, this slight bias does not disturb the consistency of our findings, since urban scaling of adoption and distance decay of spreading have similar patterns in the full network and in the 10\% sample we apply in the ABM.

\clearpage

\section*{Supporting Information 4: Calibration of ABM parameters and their influence on adoption}
\subsection*{Fitting the ABM to the diffusion data}

We fit our basic ABM model to the diffusion data using the method of Xiao et al. \cite{xiao2017bass}. The first step in the fitting is finding the linear transformation between the macroscopic $p$ and $q$ parameters of the solution of the Bass differential equation (see Eq.\ref{Eq1}), and the microscopic $p^{ABM}$ and $q^{ABM}$ parameters that drive the neighborhood adoption in the ABM (see Eq.\ref{eq:abm}). 

\begin{equation}
\left(\begin{array}{c}p^{ABM}\\q^{ABM}\end{array}\right) = \underline{\underline{C}}\left(\begin{array}{c}p\\q\end{array}\right)+\underline{\varepsilon}
\label{eq:param_transf}
\end{equation}

First, to achieve this, we run several ABM models with all possible $(p,q)$ pairs, where $p\in\{0.25^{-4}$,$0.5^{-4}$,$0.75^{-4}$,$\dots,2^{-4}\}$ and $q\in\{0.08,0.1,0.12,\dots,0.2\}$, and fit the solution of the Bass equation with nonlinear least squares method to all of the adoption curves. Thus, we get the $(p,q)$ pairs corresponding to the $(p^{ABM},q^{ABM})$ values, and by using OLS, we can fit both $\underline{\underline{C}}$ and $\underline{\varepsilon}$.

\vskip 0.1in
Second, we fit the Bass DE solution to the empirical adoption curve using again the nonlinear least squares method. From this fit, we get $\hat{p}=0.0001570$ and $\hat{q}=0.1047$. Substituting these values into Eq.~\ref{eq:param_transf}, we get our initial estimates $p_0^{ABM}=0.0001939$ and $q_0^{ABM}=0.1191$ for the microscopic parameter values.

\vskip 0.1in
Starting out from this $(p_0,q_0)$ pair, we set up a grid in the $(p,q)$ parameter space with $\Delta p=0.00001$ and $\Delta q=0.01$. We are going to run ABMs corresponding to the $(p,q)$ pairs on this grid, and we characterize the goodness of fit of these ABMs with respect to the empirical data by calculating the sum of the squared deviation of the ABM adoption curve from the empirical adoption curve (SSE). We keep track of the already visited grid points, the SSE at each gridpoint, and the two gridpoints with the least SSEs so far. In each search step, we take these two points, and we run ABMs and calculate the corresponding SSEs for all of their neighboring gridpoints $(p\pm \Delta p, q\pm \Delta q)$ that we have not visited yet. Then, we determine the two new least SSE gridpoints, and continue the search. When there are no new neighbors for the two selected least SSE points that have not been visited yet, we stop the search, and select the parameter pair with the least SSE to be the parameters for the fitted ABM. Our final parameters after this optimization step are: $p_{opt}^{ABM} = 0.0001940, q_{opt}^{ABM} = 0.1191.$

\subsection*{Selecting parameters to control for adoption threshold distribution}

\begin{figure}[h]
    \centering
    \includegraphics[width=\linewidth]{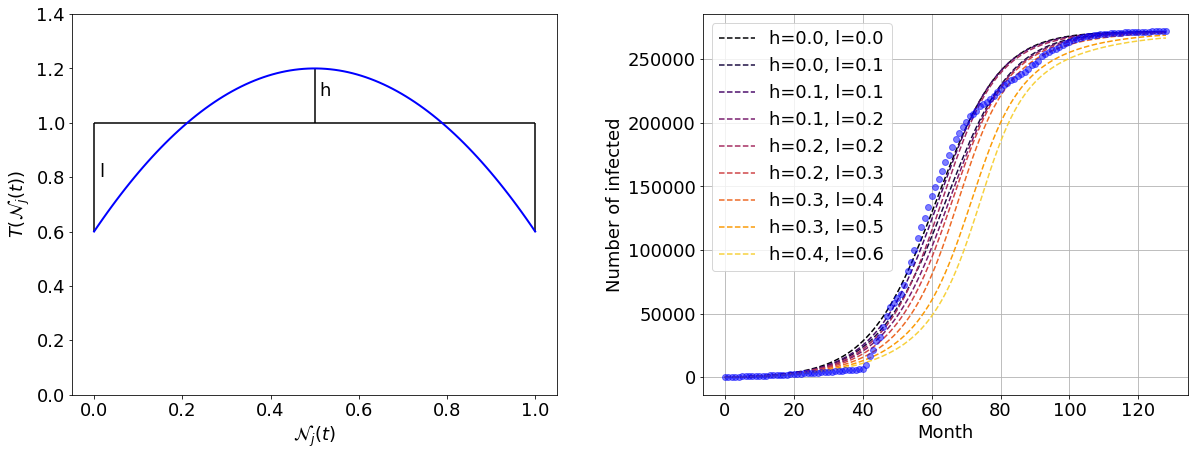}
    \caption{The transformation function for the modified ABM and parameter selection.}
    \label{fig:transform}
\end{figure}

To find the optimal values of $h$ and $l$, we run ABMs with the previously calculated $p_{opt}^{ABM}$ and $q_{opt}^{ABM}$ parameters for different $(h,l)$ parameter pairs where $h\in\{0,0.1,0.2,\dots 1\}$ and $l\in\{0,0.1,0.2,\dots 1\}$. We then select the combinations for which the error was below the threshold $\log_{10}SSE<10.2$. For these, we calculate the Pearson correlation of the peak adoption time of the largest towns (where population is greater than 5000) in the dataset. Then, as an alternative ABM model, we select $h=0.2$ and $l=0.2$, since this combination gives the highest correlation $\rho = 0.12$ apart from the original $h=0,\ l=0$ model, for which $\rho = 0.14$. Figure \ref{fig:transform} illustrates $T(x,h,l)$ from Eq. \ref{eq:t} (left) and CDF of ABM adoption considering various levels of $h$ and $l$ (right).

\subsection*{The influence of transformation function on adoption probability}
\begin{figure}[h]
    \centering
    \includegraphics[width=0.8\linewidth]{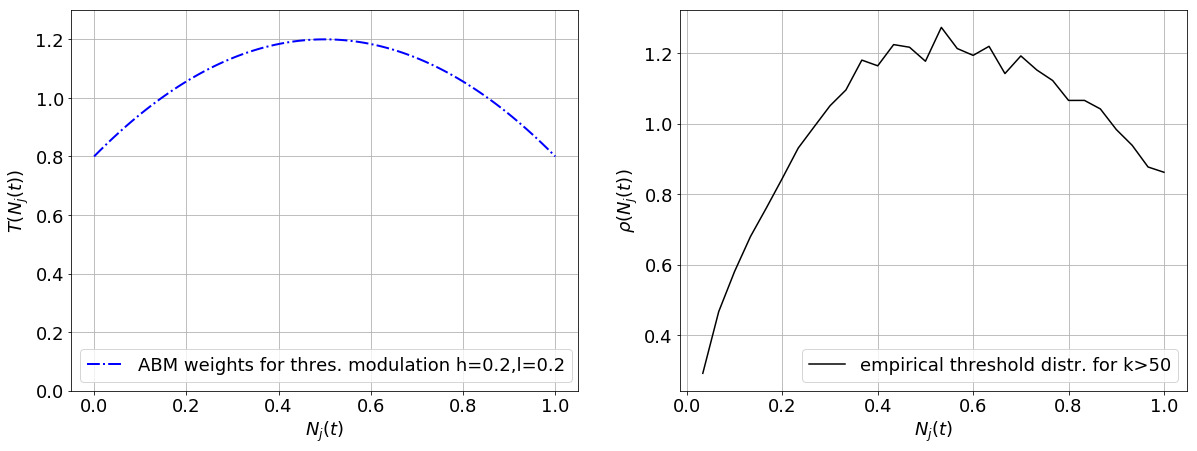}
    \caption{Threshold distribution and its modulation.}
    \label{fig:threshold_transform}
\end{figure}

Figure \ref{fig:threshold_transform} illustrates  the transformation function $T(\mathcal{N}_j(t),h=0.2,l=0.2)$ and the empirical distribution of $\mathcal{N}_j(t)$ on the full network. In this paper, we do not aim to develop a perfect $T$ to weight adoption probability that can reproduce the empirical $\mathcal{N}_j(t)$ distribution. Instead, we intend to modify adoption probability in a simple way and motivated by the threshold distribution. Our approach captures the notion that $\mathcal{N}_j(t)$ peaks between 0.4 and 0.6 (Figure \ref{fig:threshold_transform}). However, empirical $\mathcal{N}_j(t)$ are relatively rare below 0.3 (these are high degree individuals, as reported in Figure \ref{fig:fig3}B) that is not reflected by our $T$.

\begin{figure}[h]
    \centering
    \includegraphics[width=0.8\linewidth]{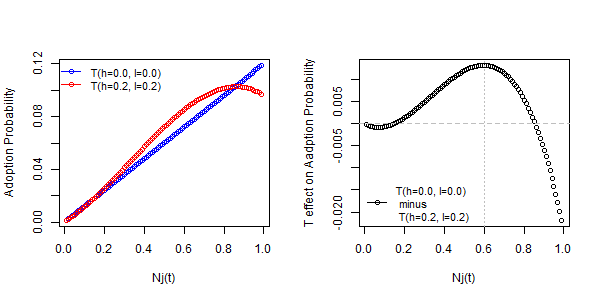}
    \caption{Linear and non-linear impact of neighbors on adoption probability.}
    \label{fig:T_effect}
\end{figure}

Substituting $T$ in Eq.\ref{eq:abm} with Eq.\ref{eq:t} gives us adoption probability at $\mathcal{N}_j(t))$, which equals  $\hat{p}^{\mbox{ABM}}+\mathcal{N}_j(t)\times \hat{q}^{\mbox{ABM}}$ in case $h=0.0$ and $l=0.0$ and $\hat{p}^{\mbox{ABM}}+(-1.6\mathcal{N}_j(t)^3 + 1.6\mathcal{N}_j(t)^2 + 0.8\mathcal{N}_j(t))\times \hat{q}^{\mbox{ABM}}$ in case $h=0.2$ and $l=0.2$. We substitute $\hat{p}^{\mbox{ABM}}$ and $\hat{q}^{\mbox{ABM}}$ values and plot adoption probabilities as a function of $\mathcal{N}_j(t))$ in Figure \ref{fig:T_effect} (left) and also their differences (right).

Setting $h=0.2$ and $l=0.2$, instead of $h=0.0$ and $l=0.0$, slightly decreases adoption probability until $\mathcal{N}_j(t))=0.2$ but provides higher probability for $\mathcal{N}_j(t))$ values between 0.2 and 0.8. The additional probability of  $h=0.2$ and $l=0.2$ is highest at $\mathcal{N}_j(t))=0.6$. Adoption probability of the $h=0.2$ and $l=0.2$ setting declines at $\mathcal{N}_j(t))>0.8$  such that probability at $\mathcal{N}_j(t))=1$ is approximately equal to the probability at $\mathcal{N}_j(t))=0.7$.

\clearpage

\section*{Supporting Information 5: Urban scaling estimates with control variables}

To understand, whether urban scaling of adoption is governed by demographic characteristics of towns, we run multiple OLS regressions with number of adopters across life-cycle stages as dependent variable. Independent variables include town population (log) and further measures that have been used in previous studies to predict adoption rate, or to investigate inequalities: development level (average income\cite{lengyel2016online}), inequalities (Gini of income\cite{toth2019inequality}), internet infrastructure and media presence (Telecom Composite Index, Number of TV, Number of School PC\cite{lengyel2016online}), physical barriers of social interaction in towns (Rail-River Division \cite{toth2019inequality}), segregation (Ethnic Entropy\cite{toth2019inequality}), town hierarchy (Subregion Centre\cite{lengyel2016online}).

\vskip 0.1in
We find a robust urban scaling coefficient reported in Figures \ref{fig:fig4}A and C. Economic development of towns measured in average salary increases adoption at all phases of the life-cycle prediction; whereas development in terms of telecommunication infrastructure facilitates adoption in the Innovation phase only.

\begin{table}[!htbp] \centering 
  \caption{Regression Results} 
  \label{} 
\begin{tabular}{@{\extracolsep{5pt}}lccc} 
\\[-1.8ex]\hline 
\hline \\[-1.8ex] 
 & \multicolumn{3}{c}{\textit{Dependent variable:}} \\ 
\cline{2-4} 
\\[-1.8ex] & \multicolumn{3}{c}{Innovator Users (log)} \\ 
\\[-1.8ex] & (1) & (2) & (3)\\ 
\hline \\[-1.8ex] 
 Population (log) & 1.342$^{***}$ & 1.297$^{***}$ & 1.083$^{***}$ \\ 
  & (1.139, 1.546) & (1.183, 1.410) & (1.047, 1.118) \\ 
  & & & \\ 
 Average salary & 0.001$^{***}$ & 0.0003$^{**}$ & 0.0001 \\ 
  & (0.0003, 0.001) & (0.0001, 0.001) & ($-$0.00002, 0.0001) \\ 
  & & & \\ 
 Gini & 0.146 & 0.236 & 0.044 \\ 
  & ($-$0.808, 1.100) & ($-$0.294, 0.766) & ($-$0.120, 0.208) \\ 
  & & & \\ 
 Telcom index & 0.082$^{*}$ & 0.041 & $-$0.016$^{**}$ \\ 
  & ($-$0.012, 0.175) & ($-$0.011, 0.093) & ($-$0.032, $-$0.0004) \\ 
  & & & \\ 
 TV use & $-$0.002 & 0.003 & 0.001 \\ 
  & ($-$0.009, 0.005) & ($-$0.001, 0.007) & ($-$0.0003, 0.002) \\ 
  & & & \\ 
 PC in school & $-$0.002 & $-$0.002 & 0.001 \\ 
  & ($-$0.007, 0.004) & ($-$0.005, 0.001) & ($-$0.0004, 0.001) \\ 
  & & & \\ 
 RRDI & 0.067 & $-$0.014 & $-$0.021 \\ 
  & ($-$0.167, 0.302) & ($-$0.147, 0.118) & ($-$0.062, 0.020) \\ 
  & & & \\ 
 Ethnic entropy & $-$0.454 & $-$0.249 & $-$0.006 \\ 
  & ($-$1.422, 0.514) & ($-$0.795, 0.298) & ($-$0.175, 0.163) \\ 
  & & & \\ 
 Town & $-$0.070 & $-$0.022 & $-$0.012 \\ 
  & ($-$0.250, 0.109) & ($-$0.120, 0.077) & ($-$0.042, 0.019) \\ 
  & & & \\ 
 Constant & $-$5.161$^{***}$ & $-$4.259$^{***}$ & $-$2.205$^{***}$ \\ 
  & ($-$6.740, $-$3.582) & ($-$5.145, $-$3.373) & ($-$2.480, $-$1.931) \\ 
  & & & \\ 
\hline \\[-1.8ex] 
Observations & 143 & 149 & 149 \\ 
R$^{2}$ & 0.726 & 0.869 & 0.978 \\ 
Adjusted R$^{2}$ & 0.658 & 0.838 & 0.973 \\ 
\hline 
\hline \\[-1.8ex] 
\textit{Note: 95\% Confidence Interval in parentheses}  & \multicolumn{3}{r}{$^{*}$p$<$0.1; $^{**}$p$<$0.05; $^{***}$p$<$0.01} \\ 
\end{tabular} 
\end{table}

\clearpage

\section*{Supporting Information 6: Estimates and confidence intervals of urban scaling coefficients in the ABM sample}

We estimate the logarithm of adopters in towns with the logarithm of town population using an ordinary least squares regression. Table \ref{tab:Scaling} details Figure \ref{fig:fig4}C by reporting 95\% confidence intervals for each estimates. All coefficients are significantly above 1. This indicates super-linear scaling meaning that adoption concentrates in large towns.

\begin{table}[!h]\centering
  \caption{Urban Scaling Coefficients} 
  \label{Scaling} 
\begin{tabular}{lccc}
\hline
                 & Innovators  & Early Adopters & Majority and Laggards \\ \hline
Data             & 1.34        & 1.26           & 1.06                  \\
                 & (1.18,1.57) & (1.16,1.36)    & (1.02,1.09)           \\ \hline
DE               & 1.19        & 1.18           & 1.04                  \\
                 & (1.13,1.25) & (1.14,1.23)    & (1.01,1.08)           \\ \hline
ABM(h=0.0,l=0.0) & 1.12        & 1.10           & 1.10                  \\
                 & (1.02,1.21) & (1.04,1.15)    & (1.06,1.13)           \\ \hline
ABM(h=0.2,l=0.2) & 1.22        & 1.13           & 1.08                  \\
                 & (1.11,1.34) & (1.05,1.21)    & (1.05,1.12)          \\ \hline
\end{tabular}
\label{tab:Scaling}
\end{table}

\vskip 0.1in

\section*{Supporting Information 7: Assortativity of adoption fuels peak prediction bias in large towns}

\begin{figure}[!b]
    \centering
    \includegraphics[width=0.8\linewidth]{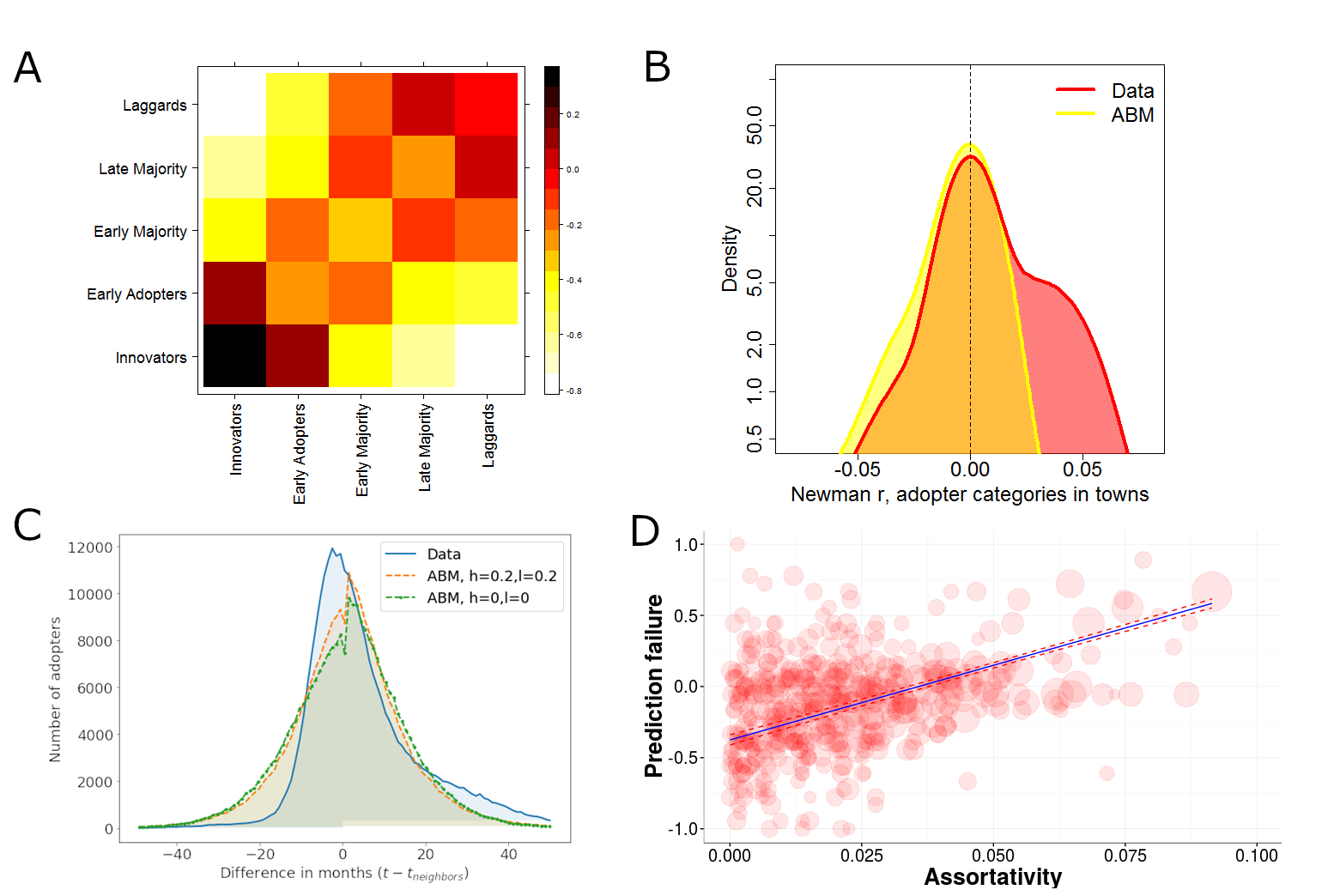}
    \caption{Assortativity in adoption and its bias in peak prediction. 
    \textbf{A.} Assortative mixing of adoption categories.  
    \textbf{B.} Assortative mixing (measured by Newman’s r) is higher in the empirical data than in the ABM(h=0.0, l=0.0). 
    \textbf{C.} Difference of adoption time (measured in months) between the user and his/her direct
    connections is smaller in the empirical data than in both versions of the ABM.
    \textbf{D.}  Assortative mixing of adopter categories by Rogers correlates with the standardized Prediction Error in the town. Size of the dots denotes the log of town population; blue solid line represents predicted values from a linear regression with 95\% confidence interval.}
    \label{fig:SI8}
\end{figure}

Connections of individuals with similar tendency to adopt, or assortative mixing, is crucial in spatial spreading. However, predicting the likelihood of adoption is the aim of diffusion models and a priori labeling of individuals in these models would be a paradox.

\vskip 0.1in
To illustrate adoption assortativity in our data in Figure \ref{fig:SI8}A, we calculated the number of links between groups \textit{W\textsubscript{ij}} and compared it to the expected number of ties \textit{E(W\textsubscript{ij})} for which uniform distribution of links across the groups is assumed and is calculated by $\frac{\sum_{j} W_i * \sum_{i} W_j}{\sum_{}W_ij} $. We have transformed the $\frac{W_ij}{E(W_ij)} $ ratio into the (-1; 1) interval using the $\frac{x-1}{x+1}$ formula. This indicator is positive if the observed number of ties exceeds the expected number of ties and negative otherwise. The plot suggests that assortative mixing fragment the network into categories of Innovators and Early Adopters who are only loosely connected to Late Majority and Laggard users.

\vskip 0.1in
To characterize assortativity on the town level, we classified each user into the adopter categories stated by Rogers\cite{rogers2010diffusion} and calculated Newman's assortativity \textit{r} \cite{newman2003mixing} for every town. This indicator takes the value of 0 when there is no assortative mixing by adopter types and a positive value when links between identical adopter types are more frequent than links between different adopter types. Figure \ref{fig:SI8}B demonstrates the similarity of peers in each town using the Newman $r$ index of assortative mixing \cite{newman2003mixing}. In many towns, the empirical data has a stronger assortativity than the ABM(h=0.0, l=0.0). This phenomenon is due to adoption time lag differences depicted in Figure \ref{fig:SI8}C. Here we contrast ABM(h=0.0, l=0.0) and ABM(h=0.2, l=0.2) with empirical data in terms of the average difference between adoption time between each ego and the time of adoption of his/her network neighbors. The ABM differs from the empirical data in determining how fast individuals follow their connections. These observations confirm that assortative mixing in terms of adoption tendency is an important feature of spatial diffusion of innovation.

To test how assortative mixing influences the spatial prediction of the diffusion ABM(h=0.0, l=0.0) in Figure \ref{fig:SI8}D, we estimated the prediction error with Newman's \textit{r} with ordinary least square estimator and used the number of OSN users in the town as weights in the regression. The ABM predicted adoption earlier in the majority of small towns, where no assortative mixing was found. On the contrary, the ABM predicted adoption late in large towns, where Innovators and Early Adopters were only loosely connected to Early- and Late Majority and Laggards. In case, we do not include weights in the regression, the point estimate of assortativity is not significant. These findings confirm that assortativity in terms of the adoption probability influences diffusion \cite{watts2002simple,toole2012modeling} and fuels peak prediction bias in large towns.

\vskip 0.1in

\section*{Supporting Information 8: Confidence intervals of Prediction Error estimations}

We estimate the Prediction Error of DE, ABM(h=0.0, l=0.0) and ABM(h=0.2,l=0.2) models with town-level social network variables and geographical characteristics using ordinary least squares regressions. Table \ref{tab:error_est} details Figure \ref{fig:fig5}C by reporting 95\% confidence intervals for each estimates.

\begin{table}[!h]\centering
  \caption{Prediction Error Estimates} 
  \label{} 
\begin{tabular}{lccc}
\hline
                 & DE  & ABM(h=0.0,l=0.0) & ABM(h=0.0,l=0.0) \\ \hline
Distance from Capital            &    0.035    &    -0.079        &  -0.100               \\
                  & (0.021,0.048) & (-0.109,-0.048)    & (-0.134,-0.065)           \\ \hline
N. of Users              &    -0.011    &    0.020       &    0.023              \\
                 & (-0.015,-0.006) & (0.010,0.030)    & (0.012,0.034)           \\ \hline
Avg. Path Length &  -0.017       &    0.026        &    0.026               \\
                 & (-0.024,-0.011) & (0.013,0.038)   &    (0.012,0.040)        \\ \hline
Modularity &    -0.065     &    0.074     &   0.098               \\
                 & (-0.089,-0.040) & (0.020,0.127)    & (0.037,0.158)          \\ \hline
Transitivity &   -0.012      &   -0.041      &    0.009              \\
                 & (-0.029,0.005) & (-0.076,-0.006)    & (-0.029,0.049)          \\ \hline
Density &   0.021      &  -0.051       &          -0.007        \\
                 & (0.007,0.035) & (-0.079,-0.023)    & (-0.038,0.024)          \\ \hline
\end{tabular}
\label{tab:error_est}
\end{table}

\clearpage

\section*{Supporting Information 9: Regression table for ABM Prediction Error}

To understand, whether Prediction Error of adoption peaks is governed by demographic characteristics of towns, we run multiple OLS regressions with Prediction Error of DE and ABM predictions as dependent variable. Independent variables include geographical variables that we focus on (population and distance) and further measures that have been used in previous studies to predict adoption rate, or to investigate inequalities: development level (average income\cite{lengyel2016online}), inequalities (Gini of income\cite{toth2019inequality}), internet infrastructure and media presence (Telecom Composite Index, Number of TV, Number of School PC\cite{lengyel2016online}), physical barriers of social interaction in towns (Rail-River Division \cite{toth2019inequality}), segregation (Ethnic Entropy\cite{toth2019inequality}), town hierarchy (Subregion Centre\cite{lengyel2016online}).

\vskip 0.1in
We find that Population and Distance influence ABM Prediction Error as reported in the main text. Also, prediction is slightly late in towns that are relatively  developed (measured by average income). The rest of the socio-economic variables, however, do not have significant point estimates.

\begin{table}[!htbp] \centering 
  \caption{Regression Results} 
  \label{} 
\begin{tabular}{@{\extracolsep{5pt}}lccc} 
\\[-1.8ex]\hline 
\hline \\[-1.8ex] 
 & \multicolumn{3}{c}{\textit{Dependent variable:}} \\ 
\cline{2-4} 
\\[-1.8ex] & Pred. Fail., DE & Pred. Fail., ABM(h=0.0, l=0.0) & Pred. Fail., ABM(h=0.2, l=0.2) \\ 
\\[-1.8ex] & (1) & (2) & (3)\\ 
\hline \\[-1.8ex] 
 Population (log) & 0.039$^{***}$ & 0.030$^{***}$ & 0.038$^{***}$ \\ 
  & (0.023, 0.054) & (0.013, 0.047) & (0.019, 0.057) \\ 
 Distance from Budapest, km (log) & $-$0.021 & $-$0.064$^{***}$ & $-$0.076$^{***}$ \\ 
  & ($-$0.053, 0.011) & ($-$0.098, $-$0.029) & ($-$0.114, $-$0.038) \\ 
 Average salary & $-$0.00002 & 0.0001 & 0.0001$^{***}$ \\ 
  & ($-$0.0001, 0.00004) & ($-$0.00001, 0.0001) & (0.00005, 0.0002) \\ 
 Gini & 0.056 & 0.042 & 0.067 \\ 
  & ($-$0.046, 0.159) & ($-$0.069, 0.153) & ($-$0.056, 0.190) \\ 
 Telcom index & $-$0.001 & 0.004 & $-$0.003 \\ 
  & ($-$0.013, 0.011) & ($-$0.009, 0.016) & ($-$0.017, 0.011) \\ 
 TV use & $-$0.0002 & 0.0004 & 0.0001 \\ 
  & ($-$0.001, 0.001) & ($-$0.0004, 0.001) & ($-$0.001, 0.001) \\ 
 PC in school & 0.0002 & 0.00002 & 0.0003 \\ 
  & ($-$0.0005, 0.001) & ($-$0.001, 0.001) & ($-$0.001, 0.001) \\ 
 RRDI & 0.006 & $-$0.013 & 0.011 \\ 
  & ($-$0.023, 0.036) & ($-$0.044, 0.019) & ($-$0.024, 0.046) \\ 
 Ethnic entropy & $-$0.084 & $-$0.112$^{*}$ & $-$0.013 \\ 
  & ($-$0.197, 0.028) & ($-$0.233, 0.010) & ($-$0.147, 0.122) \\ 
 Town & 0.013 & 0.008 & $-$0.002 \\ 
  & ($-$0.012, 0.039) & ($-$0.019, 0.036) & ($-$0.033, 0.028) \\ 
 Constant & 0.075 & $-$0.029 & $-$0.117 \\ 
  & ($-$0.295, 0.445) & ($-$0.431, 0.372) & ($-$0.561, 0.327) \\ 
\hline \\[-1.8ex] 
Observations & 2,237 & 2,237 & 2,237 \\ 
R$^{2}$ & 0.026 & 0.030 & 0.038 \\ 
Adjusted R$^{2}$ & 0.014 & 0.017 & 0.025 \\ 
\hline 
\hline \\[-1.8ex] 
\textit{Note:95\% CI in parentheses.}  & \multicolumn{3}{r}{$^{*}$p$<$0.1; $^{**}$p$<$0.05; $^{***}$p$<$0.01} \\ 
\end{tabular} 
\end{table}

\end{document}